\documentclass[11pt]{article}
\usepackage{mathptmx}
\newtheorem{thm}{Theorem}[section]
\newtheorem{proposition}[thm]{Proposition}

\newtheorem{lemma}[thm]{Lemma}

\begin{document}

\begin{center}

{\bf\Large On Darboux-Treibich-Verdier potentials}

\bigskip

{\bf  A. P. Veselov}
\end{center}

\medskip

\noindent Department of Mathematical Sciences,
Loughborough University,\,
Loughborough,  LE11 3TU, UK
and Moscow State University, Moscow, 119899, Russia

\medskip

\noindent E-mail address: 
A.P.Veselov@lboro.ac.uk

\medskip

{\begin{center}\normalsize\it
     To V.B. Matveev
     on his 65-th birthday
   \end{center}}

\begin{abstract}
It is shown that the four-parameter family of elliptic functions 
$$u_D(z)=m_0(m_0+1)\wp(z)+\sum_{i=1}^3 m_i(m_i+1)\wp(z-\omega_i)$$
introduced by Darboux and rediscovered a hundred years later by Treibich and Verdier, is the most general meromorphic family containing infinitely many finite-gap potentials.
\end{abstract}



\vspace{1cm}

\section{Introduction}

In 1882 in a short Comptes Rendus article Darboux \cite{D} introduced and studied the following differential equation as a generalisation of 
the famous Lam\`e equation:
$$
-\frac{d^2y}{dz^2} +(\frac{m_0(m_0+1)}{sn^2z}+\frac{m_1(m_1+1)dn^2z}{cn^2z}+\frac{m_2(m_2+1)k^2cn^2z}{dn^2z}
$$
\begin{equation}
\label{DTV}
\noindent +m_3(m_3+1)k^2 sn^2z)y=\lambda y,
\end{equation}
where $m_i,\, i=0,1,2,3$ are 4 integer parameters and $sn, cn, dn$ are the standard Jacobi elliptic functions with parameter $k$ (see \cite{WW}).
In a more convenient Weierstrass form the family of the corresponding potentials is 
\begin{equation}
\label{DTVW}
u_D(z)=m_0(m_0+1)\wp(z)+\sum_{i=1}^3 m_i(m_i+1)\wp(z-\omega_i),
\end{equation}
where $\omega_1, \omega_2, \omega_3$ are the half-periods of the corresponding elliptic curve \cite{WW}.

A hundred years later this family was rediscovered by Treibich and Verdier \cite{TV,V} in the context of the finite-gap theory. Almost at the same time Inozemtsev \cite{I} considered it in relation with the generalisations of the elliptic Calogero-Moser system.

The family ({\ref{DTVW}) was known as Treibich-Verdier potentials until about 10 years ago V.B. Matveev pointed out that it was already in the old Darboux note \cite{D}. Following \cite{MS} we will call it {\it Darboux-Treibich-Verdier} (DTV) family.

The aim of this note is to show that this family is actually a very special one: this is the most general linear family of meromorphic functions, 
which contains infinitely many finite-gap potentials (see a precise formulation below). 

I did not see this claim in the literature but I would not be surprised if in some form it was known to the experts. I should say that the proof simply follows from the modern finite-gap theory in combination with the classical theory of differential equations in the complex domain.  

This note was written on the occasion of 65-th birthday of Vladimir Borisovich Matveev, who was the first to appreciate and to demonstrate the importance of Darboux ideas for the modern theory of integrable systems.

\section{Claim and a proof}

In 1974 S.P. Novikov \cite{N} made a remarkable discovery: he showed that a real periodic 
Schr\"odinger operator $$L=-D^2+u(z),\, D=\frac{d}{dz},$$ having commuting differential operator
$$A=D^{2n+1}+a_1(z)D^{2n-1}+\dots+a_{2n}(z)$$ of an odd order $2n+1,$ has at most $n$ gaps in its spectrum (see also Lax \cite{L}). Dubrovin \cite{Du} showed later that the converse is also true: for any such operator $L$ there exists an odd-order differential operator $A$ such that 
\begin{equation}
\label{Comm}
[L,A]=0.
\end{equation}
Slightly abusing terminology, we will call any Schr\"odinger operator $L$ (not necessary real and in general with singularities) with this property and the corresponding potential $u(z)$ as {\it finite-gap.}

Note that  the commutativity equation (\ref{Comm}) is equivalent to nonlinear ordinary differential equation on $u(z)$, which is the stationary higher KdV flow \cite{N}.
Historically this equation was first investigated (without any relation to spectral theory) by Burchnall and Chaundy \cite{BC}, who proved that the operators $L$ and $A$ satisfy the algebraic relation
\begin{equation}
\label{BC}
A^2=P(L)
\end{equation}
 for some polynomial $P(x)$ of degree $2n+1.$
When the corresponding curve is non-singular the potential can be expressed explicitly in the hyperelliptic theta-functions as it was shown by Its and Matveev \cite{IM}. Further development was due to Krichever, who introduced an important general notion of Baker-Akhiezer function \cite{K}.

The following property of the finite-gap operators will play crucial role in our considerations.

\smallskip

\noindent {\it Painlev\`e property.} All the finite-gap potentials $u(z)$ are {\it meromorphic}  in the whole complex plane. The same is true for all solutions $\psi(z)$ of the corresponding Schr\"odinger equation
$$-\psi''+u(z)\psi = \lambda \psi$$
for all $\lambda.$ In generic situation this follows immediately from Its-Matveev formulas \cite{IM}, the general case see in Segal and Wilson \cite{SW}.

\smallskip

Let $f_0(z), \dots, f_N(z), \, z\in \mathbf C$ be some meromorphic functions and
 $V$ be finite-dimensional linear subspace spanned by them: 
  \begin{equation}
\label{F}
V=\{u(z)= \sum_{i=0}^N \alpha_i f_i(z), \, \alpha_i \in \mathbf C.\}
\end{equation}
We will call such a subspace {\it finite-gap} if $V$ contains an infinite subset $K \subset V$ consisting of finite-gap potentials, such that $K$ is dense in $V$ in Zariski topology. This means that if a polynomial is vanishing at the points of $K$ then it must vanish identically. We also say that $V$ is {\it maximal} if $V$ is not contained in a larger finite-gap family.

Let $\wp(z)$ be the classical Weierstrass elliptic function with periods $2\omega_1, 2\omega_2$
and $\omega_3=\omega_1+\omega_2$ (see \cite{WW}).

\begin{thm} \label{Th} Modulo shift in $z$ there are only 3 maximal finite-gap subspaces, consisting of the following elliptic, trigonometric and rational potentials respectively:
  \begin{equation}
\label{ell}
u_{ell}(z)=\sum_{i=1}^3 \alpha_i \wp(z-\omega_i)+\alpha_4\wp(z)+\alpha_0,
\end{equation} 
\begin{equation}
\label{trig}
u_{trig}(z)=\alpha_1 \frac{a^2}{\sin^2 az} + \alpha_2 \frac{a^2}{\cos^2 az} +\alpha_0,
\end{equation} 
\begin{equation}
\label{rat}
u_{rat}(z)= \alpha_1 \frac{1}{z^2} + \alpha_0.
\end{equation}
\end{thm}

The fact that the potentials $u_{ell}(z)$ are finite-gap for parameters from the set
$$K=\{\alpha_i=m_i(m_i+1), \, m_i \in \mathbf Z, i=1,\dots, 4, \alpha_0 \in \mathbf C\}$$ is due to Treibich-Verdier \cite{TV}.
Note that the set $K$
is indeed dense in $V \approx \mathbf C^5$ in Zariski topology.
The trigonometric potentials 
$$u_{trig}(z)= \frac{m_1(m_1+1)a^2}{\sin^2 a z} + \frac{m_2(m_2+1)a^2}{\cos^2 az}$$
with integer $m_1, m_2$ sometimes are called {\it P\"oschl-Teller} potentials, although they were studied already by Darboux \cite{D2}. They are known to be the result of Darboux transformations applied to $u=0$ (see e.g. \cite{GM}) and thus are finite-gap.
The same is true about rational case when $\alpha_1=m(m+1)$ with integer $m.$
Note that trigonometric and rational families can be considered as a limits of the elliptic family when one or both of the periods go to infinity. 

Thus we must show only that there are no more maximal finite-gap subspaces. To prove this we need the following crucial fact, showing the relation of this question with groups generated by reflections.

Let $V$ be a finite-gap subspace. Consider its singular set $\Sigma=\Sigma(V)$, which is the union of the singularity sets (poles) of the corresponding functions $f_0, \dots, f_N$. Let $z_0 \in \Sigma$ be any such pole and consider the symmetry $s$ with respect to $z_0:$
$s(z)=2z_0-z.$ We claim that $s(\Sigma)=\Sigma$ for any $z_0 \in \Sigma.$

\begin{proposition}
\label{Pr}
The singular set $\Sigma(V)$ of any finite-gap subspace $V$ is symmetric with respect to each of its points.
Moreover, every potential in $V$ is invariant under such a symmetry.
\end{proposition}

To prove this we use the Painlev\`e property of the finite-gap operators and Frobenius analysis of the corresponding Schr\"odinger equation
$-\psi''+u(z)\psi = \lambda \psi$ in the complex domain.

Without loss of generality we can assume that $f_0=1$ and $z_0=0$ is the pole of $f_1(z)$ (all other $f_i$ can be assumed regular at $0$).
It is known that the finite-gap potentials have only second-order poles with zero residues (see e.g. \cite{SW}), so the same must be true for the functions $f_1, \dots, f_N.$ The Laurent expansion at zero of the potentials from $V$ has the form
  \begin{equation}
\label{c}
u(z)=\frac{c_{-2}}{z^2}+c_0+c_1 z+c_2z^2+\dots
\end{equation}
with $c_{-2}=\alpha_1$ and other $c_i=c_i(\alpha_0,\dots, \alpha_N)$ are certain linear functions of the parameters.
Let us consider the corresponding Schr\"odinger equation
\begin{equation}
\label{ode}
-\varphi''+u(z)\varphi = \lambda \varphi
\end{equation}
and ask when it has all solutions meromorphic in the vicinity of $z=0$ for all $\lambda.$
In that case we will say that (\ref{ode}) has {\it trivial monodromy} around $z=0.$

The following important lemma looks classical, but as far as I know first appeared in Duistermaat and Gr\"unbaum \cite{DG}.

\begin{lemma} \cite{DG}.
The equation (\ref{ode}) has trivial monodromy around $z=0$ if and only if the Laurent coefficients of the corresponding potential (\ref{c})  satisfy the conditions:
\begin{equation}
\label{loc1}
c_{-2} = m(m+1), \, m\in {\bf Z}_+
\end{equation}
and \begin{equation}
\label{loc2}
c_{2k-1} = 0, \, k = 1,..., m.
\end{equation}
\end{lemma}

The proof follows the classical Frobenius line:
the solutions must be meromorphic, so
$$ \varphi = z^{-\mu}(1 + \sum_{i=1}^{\infty} \xi_i z^{i}).$$
Substituting this into equation and equating the coefficents we see that $\mu$ must satisfy the characteristic equation
$\mu(\mu + 1) = c_{-2},$ which implies (\ref{loc1}). This is not enough yet since $\varphi$
may have a logarithmic term. A simple analysis  \cite{DG} shows
that the logarithmic terms are absent for all $\lambda$ if and only if
in addition to (\ref{loc1}) we have (\ref{loc2}).

Now we can prove the proposition. Since finite-gap potentials have Painlev\`e property they have trivial monodromy. Thus we have infinitely many potentials (\ref{c}) with parameters $c_{-2}=m(m+1)$ for infinitely many integers $m.$ This means that we have all odd coefficients $c_{2k-1}(\alpha)=0$ for the corresponding $\alpha=(\alpha_0, \dots, \alpha_N) \in K.$ Now because $K$ is Zariski dense this implies that $c_{2k-1}=0$ for all potentials from $V$. This means that all the potentials from $V$ are even functions and hence their singularity set is symmetric with respect to $z_0=0,$ which completes the proof.

\smallskip

Now we need the following elementary geometric
\begin{lemma}
Suppose that a discrete set $\Sigma\subset \mathbf C$ is invariant under reflection $z \rightarrow 2z_0-z$
with respect to every point $z_0 \in \Sigma.$ Then there are only 3 possibilities:
  \begin{equation}
\label{ell1}
\Sigma=\{z=z_0+k_1\omega_1+k_2 \omega_2, \, k_1, k_2 \in \mathbf Z, \, \omega_1/\omega_2 \notin \mathbf R\},
\end{equation}
  \begin{equation}
\label{trig1}
\Sigma=\{z=z_0+k\omega, \, k \in \mathbf Z, \, \omega\neq 0\},
\end{equation}
\begin{equation}
\label{rat1}
\Sigma=\{z=z_0\}.
\end{equation}
\end{lemma}
The proof is simple: if we have two different points $z_0,z_1 \in \Sigma$ then the composition of the symmetries with respect to them gives a shift $z \rightarrow z + 2\omega,\, \omega=z_1-z_0.$
This means that the set must contain the one-dimensional lattice $\{z=z_0+k\omega, \, k \in \mathbf Z\}.$ If we have more points then it is easy to see that we have either larger one-dimensional or two-dimensional lattice. The only remaining case is a one-point set.

To derive our main result consider first the case (\ref{ell1}) when $\Sigma$ is a shifted two-dimensional lattice. The group generated by reflections with respect to the corresponding points acts on $\Sigma$ and has 4 orbits corresponding to the different parities of $k_1$ and $k_2.$ The corresponding meromorphic functions are elliptic and must have the Darboux-Treibich-Verdier form (\ref{ell}) shifted by $z_0$. In the case (\ref{trig1})
the potentials are trigonometric functions and according to Airault, McKean and Moser \cite{AMM} must be of the general form
$$u(z) = \sum_{i=1}^N \frac{m_i(m_i+1)a^2}{\sin^2a(z-z_i)}+const, \,\, a=\pi/\omega.$$
In order to fit (\ref{trig1}) up to a shift in $z$ they must be of the form (\ref{trig}). The remaining rational case is similar. This completes the proof of Theorem \ref{Th}.

{\it Remark.} As we can see from the proof Zariski density can be replaced by the condition that $K$ is not contained in any hyperplane in $V.$

\section{Concluding remarks}

What we saw here is another demonstration of a deep link between finite-gap property and trivial monodromy in the complex domain. The last property was extensively studied in the end of XIX-th century starting from the work by Hermite and Picard's work in 1870s. 
Darboux work \cite{D} also appeared in this context.
The famous S. Kowalevskaya's work on integrable case in rigid body dynamics was further development of this idea. A link with Kowalevskaya's work had been emphasized already at a very early stage of the modern finite-gap theory by Dubrovin, Matveev and Novikov in \cite{DMN}, the relation of Picard's work with elliptic finite-gap potentials  was studied in detail by Gesztesy and Weikard \cite{GW}.

I should mention that there are Schr\"odinger operators with trivial monodromy, which are not finite-gap.
A simple example is given by $$u(z)=\frac{m(m+1)}{z^2}+P(z)$$ for any even polynomial $P(z).$
For more interesting examples, including Painlev\`e-IV transcendents, we refer to \cite{GV,HV,Ves}. 
Note that as it follows from the proof Proposition \ref{Pr} is still valid if we replace finite-gapness by trivial monodromy property.

An obvious question is about multi-dimensional analogue of our result. It is natural to expect that the answer would be related to the reflection groups and should be given by the corresponding {\it generalised quantum Calogero-Moser systems}, see the elliptic case in \cite{I, CEO}. I would like to note also that the multi-dimensional trivial monodromy condition plays a key role in the theory of Huygens' principle \cite{CFV}.

\end{document}